\documentclass[twocolumn,epjc3]{svjour3}  
\RequirePackage{graphicx}
\RequirePackage{latexsym}
\RequirePackage[numbers,sort&compress]{natbib}
\RequirePackage[colorlinks,citecolor=blue,urlcolor=blue,linkcolor=blue]{hyperref}
\RequirePackage{xcolor}
\RequirePackage{subcaption}
\RequirePackage{amsmath}
\RequirePackage{fix-cm}
\RequirePackage{xspace}
\newcommand{\code}[1]{\textsc{#1}}
\newcommand{\LP}{LP$_\text{CNN}$\xspace}

\journalname{Eur. Phys. J. C}
\begin{document}

\title{Tagging the Higgs boson decay to bottom quarks with colour-sensitive observables and the Lund jet plane}

\author{%
  Luca~Cavallini\thanksref{paviaUNI,lc} 
  \and
  Andrea~Coccaro\thanksref{genovaINFN,ac}
  \and
  Charanjit~K.~Khosa\thanksref{BristolUni,genovaUNI,genovaINFN,ck}
  \and
  Giulia~Manco\thanksref{paviaUNI,paviaINFN,gm}
  \and
  Simone~Marzani\thanksref{genovaUNI,genovaINFN,sm} 
  \and
  Fabrizio~Parodi\thanksref{genovaUNI,genovaINFN,fp}
  \and
  Daniela~Rebuzzi\thanksref{paviaUNI,paviaINFN,dr}
  \and
  Alberto~Rescia\thanksref{paviaUNI,ar}
  \and
  Giovanni~Stagnitto\thanksref{zurich,gs}
}

\thankstext{lc}{e-mail: luca.cavallini02@universitadipavia.it}
\thankstext{ac}{e-mail: andrea.coccaro@ge.infn.it}
\thankstext{ck}{e-mail: charanjit.kaur@bristol.ac.uk}
\thankstext{gm}{e-mail: giulia.manco@pv.infn.it}
\thankstext{sm}{e-mail: simone.marzani@ge.infn.it}
\thankstext{fp}{e-mail: fabrizio.parodi@ge.infn.it}
\thankstext{dr}{e-mail: daniela.rebuzzi@unipv.it}
\thankstext{ar}{e-mail: alberto.rescia@desy.de}
\thankstext{gs}{e-mail: giovanni.stagnitto@physik.uzh.ch}

\institute{%
  Dipartimento di Fisica, Universit\`a di Pavia, Via A. Bassi 6, 27100 Pavia, Italy
  \label{paviaUNI}
  \and
  INFN, Sezione di Genova, Via Dodecaneso 33, 16146 Genova, Italy
  \label{genovaINFN}
  \and
  H.H. Wills Physics Laboratory, University of Bristol, Tyndall Avenue, Bristol BS8 1TL, UK
  \label{BristolUni}
  \and
  Dipartimento di Fisica, Universit\`a di Genova, Via Dodecaneso 33, 16146 Genova, Italy
  \label{genovaUNI}
  \and
  INFN, Sezione di Pavia, Universit\`a di Pavia, Via A. Bassi 6, 27100 Pavia, Italy  
  \label{paviaINFN}
  \and
  Physik-Institut, Universit\"{a}t Z\"{u}rich, Wintherturerstrasse 190, CH-8057 Z\"{u}rich, Switzerland
  \label{zurich}  
}

\date{ZU-TH 50/21}

\maketitle

\begin{abstract}
 We study the problem of distinguishing $b$-jets stemming from the decay of a colour singlet, such as the Higgs boson, from those originating from the abundant QCD background. 
 In particular, as a case study, we focus on associate production of a vector boson and a Higgs boson decaying into a pair of $b$-jets, which has been recently observed at the LHC.
 We consider the combination of several theory-driven observables proposed in the literature, together with Lund jet plane images, in order to design an original $Hbb$ tagger. The observables are combined by means of standard machine learning algorithms, which are trained on events obtained with fast detector simulation techniques.
 We find that the combination of high-level single-variable observables with the Lund jet plane provides an excellent discrimination performance.
 We also study the dependence of the tagger on the invariant mass of the decaying particles, in order to assess the extension to a generic $Xbb$ tagger.
\end{abstract}


\section{Introduction}
\label{sec:intro}

Since the discovery of the Higgs boson at the Large Hadron Collider (LHC) by the ATLAS~\cite{1207.7214} and CMS~\cite{1207.7235} experiments in 2012, our understanding of the properties of this particle has progressively evolved. In addition to the ``golden'' decay modes of the Higgs boson, $H \rightarrow \gamma\gamma$ and $H \rightarrow 4l$, in the past few years other decay channels have been observed, usually in association with particular production modes. 
For instance, due to its large branching ratio, the $H\rightarrow b\overline{b}$ decay plays a central role in studies that aim at probing the structure of the Higgs couplings to the fermions.
In this regard, one of the most interesting processes is the associated production of a Higgs boson $H$ and a vector boson $V$ ($W$ or $Z$), with the vector boson decaying leptonically and the Higgs boson decaying hadronically into a pair of $b$-quarks, $V(l\bar{l})H(b\bar{b})$: the decay products of the vector boson provide us with a clean experimental signature, as well as a recoil system for the Higgs particle. 
Both ATLAS and CMS experiments have reported the observation of the $H\rightarrow b\overline{b}$ decay and of the $VH$ production mode~\cite{1808.08238, 1808.08242} and the ATLAS experiment reported the first cross-section measurements~\cite{1903.04618, ATLAS:2019yhn, ATLAS:2020fcp, ATLAS:2020jwz} targeting different regimes of reconstructed transverse momenta of the vector boson and in fiducial volumes, as defined by the simplified template cross-section framework~\cite{Berger:2019wnu}. The experimental focus is therefore shifting towards precision measurements of the kinematics of the $H\rightarrow b\overline{b}$ decay channel and, as suggested in Ref.~\cite{Brehmer:2019gmn}, additional differential information, and hence discrimination of this process against sources of backgrounds, is crucial for the sensitivity to beyond-the-Standard-Model operators.
After the fragmentation and hadronisation process, the hard $b$-quarks produced by the Higgs boson decay are usually detected as two separate $b$-jets~\cite{ATLAS:2019bwq,CMS:2017wtu}.
In simulation, a $b$-jet is defined by a suitable particle-level observable, based on the angular distance of $B$-hadrons with respect to the jet axis, or by ghost association~\cite{Cacciari:2008gn,ATLAS:2013bqs}. On real data, $b$-jets are identified by means of dedicated $b$-tagging algorithms.

In order to make the most out of the large set of accumulated data,
strategies to better discriminate the $H(b\bar{b})$ process over the large
QCD background (where the pair of $b$-quarks is produced by pure strong interaction, mostly by $g(b\bar{b})$ collinear splitting) are being actively developed.
The signal/background discrimination is especially compelling in the boosted regime, when the transverse momentum of the jets is much greater than their invariant mass: in such a situation, the $b$-quarks may be close in angle, and hence reconstructed as a single jet. 
Since the seminal work of Ref.~\cite{Butterworth:2008iy}, several jet substructure techniques--- which aim to improve the discrimination performance by finding hard prongs inside a large radius jet --- have been designed, tested and implemented in the analyses by the experimental collaborations (see, for instance Ref.~\cite{Marzani:2019hun} and references therein). 
%
Broadly speaking, two main strategies exist: design high-level theoretically-motivated observables, sensitive to particular features of the signal distribution, which can be measured on data and used as single-variable discriminant; or produce some low-level representation of the jets (list of particles, calorimetric images,\dots), to be used as input for machine learning (ML) techniques.
We refer the reader to the recent literature about ML based approaches for $H \rightarrow b \bar b$ tagging~\cite{Datta:2019ndh,Lin:2018cin,Moreno:2019neq,Chakraborty:2019imr,Sirunyan:2020lcu,Chung:2020ysf,Tannenwald:2020mhq,guo2020boosted,Abbas:2020khd,Jang:2021eph,Khosa:2021cyk}, which is continuously being updated in Ref.~\cite{hepmllivingreview}.

Our case of interest is particularly challenging because both signal and background feature a similar flavoured two-prong structure. However, the two processes have a different behaviour with respect to the QCD radiation pattern. Namely, in the signal case, the $b$-jets originate from the decay of a colour singlet, and thus radiation will be mostly contained within the two $b$-quark system. Instead, in the background case, we expect QCD radiation to be more diffuse, due to colour connections with the rest of the event. Therefore, we would like to exploit observables that are particularly sensitive to the colour flows in the event. 
In this paper we select several theory-driven single-valued observables (see Sect.~\ref{sec:colobs}) and a theoretically-motivated representation of a jet (specifically, the Lund jet plane~\cite{Dreyer:2018nbf}) and build a combined $Hbb$ tagger, with the aim to exploit the best of both strategies. Such a combination is performed by means of standard ML algorithms, for different input choices. We use boosted decision trees (BDT) for single-valued observables and convolutional neural networks (CNN) for Lund jet images. BDTs have been part of HEP analyses for a long time~\cite{Yang:2005nz}. CNNs are showing promising potential for image based data sets for various applications in HEP see e.g.\ \cite{Khosa:2020qrz, Khosa:2019qgp, Khosa:2019kxd}.
Moreover, in our analysis we account for the experimental detection and reconstruction of the physical quantities of interest by performing a fast detector simulation on the generated Monte Carlo events, and we assess the impact of these so-called detector effects on individual variable distributions and on the overall performance of the tagger. 
Finally, in an ideal scenario we would like to apply the same tagger for the decay products of a generic colour singlet $X$, without any prior knowledge on the value of its mass, so as to design a global $Xbb$ tagger. 
In this view, it would be desirable to keep the tagger uncorrelated with the invariant mass of the $b$-quark pair, in order to ease its calibration on $Z$+jet events and to simplify the determination of the non-resonant background shape in data-driven approaches.
 
The paper is organised as follows.
In Sect.~\ref{sec:colobs} we briefly introduce the colour sensitive observables under study and the Lund jet plane.
In Sect.~\ref{sec:analysis} we discuss the event generation set-up and the selection
cuts adopted in our analysis.
In Sect.~\ref{sec:results} we study the individual distributions of the colour sensitive observables and the output of the Lund jet plane CNN for the signal and the background processes, before and after detector simulation. 
Moreover, we assess the discrimination performance of the combination of several observables, by also including the Lund jet plane CNN output as an additional input.
In Sect.~\ref{sec:massbias} we discuss the BDT dependence on the invariant mass of the large radius jet used in the analysis (see Sect.~\ref{sec:analysis}), to determine the mass bias of our $Hbb$ tagger.
Finally, in Sect.~\ref{sec:concl} we draw our conclusions.

\section{Description of the observables}
\label{sec:colobs}

We make use of high-level colour sensitive variables introduced in the literature in the past few years: jet pull and its projections, namely the pull angle, $\theta_p$~\cite{Gallicchio:2010sw,Larkoski:2019urm}, and the parallel and perpendicular components of the pull vector, $t_\parallel$ and $t_\perp$~\cite{Bao:2019usu, Larkoski:2019fsm}; the colour ring $\mathcal{O}$ \cite{Buckley:2020kdp}; $D_2$ \cite{Larkoski:2014gra, Moult:2016cvt} and the Lund jet plane \cite{Dreyer:2018nbf}.
We limit ourselves to a brief introduction of the relevant variables, and we refer the interested reader to the original papers.

\subsection{Jet pull}

Let us consider a hard jet $J_a$.
The {\em pull vector} $\vec{t}$ is the jet shape observable that is defined as
\begin{equation}
  \vec{t} = \frac{1}{p_{Ta}} \sum_{i \in J_a}p_{Ti}\vert \vec{r}_i \vert^2 \hat{r}_i,
\end{equation}
where $p_{T_a}$  is the transverse momentum of the jet, and the sum runs over all the jet constituents. $y$ and $\phi$ represent rapidity and azimuthal angle, and $\vec{r}_i$ is the distance vector between the jet axis and its $i$-th constituent in the $y$-$\phi$ plane
\begin{equation}
  \vec{r}_i = (y_i - y_a, \phi_i - \phi_a).
\end{equation}
The pull vector is sensitive to the different colour connections of the entire event in which the jet is formed. 
If we consider events with two hard jets (or subjets) $J_a$ and $J_b$ that originate from the decay of a colour singlet, additional QCD radiation tends to be emitted between the two jets, causing the pull vector of $J_a$ to point in the direction of $J_b$ and vice-versa. If instead the two jets originate from the decay of a colour octet, such as the gluon, then the pull vectors will instead tend to point in different directions.

In order to make these considerations more quantitative, we can introduce suitable projections of the pull vector $\vec{t}$ along two directions: the one given by the unit vector which points from the centre of $J_a$ to the centre of $J_b$
\begin{equation}
  t_\parallel = \vec{t}\cdot \hat{n}_\parallel, \quad \text{with} \quad \hat{n}_\parallel = \frac{1}{\sqrt{\Delta y^2 + \Delta \phi^2}}\left(\Delta y, \Delta \phi \right),
\end{equation}
and the other generated by the unit vector perpendicular to $\hat{n}_\parallel$
\begin{equation}
   t_\perp = \vec{t}\cdot \hat{n}_\perp, \quad \text{with} \quad \hat{n}_\perp = \frac{1}{\sqrt{\Delta y^2 + \Delta \phi^2}}\left(-\Delta \phi, \Delta y \right),
\end{equation}
where in the above equations we have introduced $\Delta y= y_a-y_b$ and $\Delta \phi= \phi_a-\phi_b$.
We also consider $\theta_{p}$, known as the pull angle, defined as
\begin{equation}
  \theta_p = \arccos \frac{t_\parallel}{\vert \vec{t} \vert}.
\end{equation}

Of all the variables built out of the jet pull vector, the pull angle has been shown to be one of the most effective discriminants of the two different colour configurations \cite{Gallicchio:2010sw}. However, the comparison between experimental measurements of the pull angle and theoretical calculations has shown that this observable is not under good theoretical control~\cite{D0:2011lzz,ATLAS:2015ytt,ATLAS:2018olo}.
This problem can be traced back to the fact that $\theta_p$ is not infra-red and collinear (IRC) safe, but only Sudakov safe~\cite{Larkoski:2019urm} (for discussions about Sudakov safety, see Refs.~\cite{Larkoski:2013paa,Larkoski:2014wba,Larkoski:2015lea}).
Instead, $t_\parallel$ and $t_\perp$ \emph{are} IRC safe observables, so it is interesting to assess whether individually or in some combination $t_\parallel$ and $t_\perp$ possess the same discriminating power of its non IRC safe counterpart. It is also interesting to study how the discriminating power of the pull vector variables is affected when combining the variables for both jets $J_a$ and $J_b$.
Thus, we will include all the three pull vector variables, for both jets $J_a$ and $J_b$, as input to the ML algorithms. Furthermore,  although $\theta_p$ is not IRC safe, it is also included.

Finally, we note that there is potential overlap in including both $t_\parallel$, $t_\perp$ and $\theta_p$ because the jet pull is only a two-component vector. However, this is not an issue, since the machine learning algorithms are trained to be robust against interdependence between input variables.

\subsection{Jet colour ring}

The jet colour ring was introduced in Ref.~\cite{Buckley:2020kdp}, as an observable that is provably optimal, in certain kinematic limits. 
The starting point of its construction is the observation that, according to the Neyman-Pearson lemma, the ratio of the matrix elements squared for the signal and the background process should be monotonic to the optimal single-variable discriminant~\cite{Neyman:1933wgr}.
When considering a decay of a colour singlet as signal and a colour octet as background, with a subsequent gluon emission in the boosted regime, and working in the soft-collinear limit approximation, the ratio simplifies to
\begin{equation}
  \frac{\vert \mathcal{M}_S \vert^2}{\vert \mathcal{M}_B \vert^2} \simeq \frac{\theta_{ak}^2 + \theta_{bk}^2}{\theta_{ab}^2}\,,
\end{equation}
where the indices $a$ and $b$ refer to the hard partons, the index $k$ to an additional (gluon) emission, and $\theta_{ij}$'s are the angles between them.
The above considerations lead to the definition of the jet colour ring
\begin{equation}
\mathcal{O}=\frac{\Delta_{ak}^2 + \Delta_{bk}^2}{\Delta_{ab}^2},
\label{color ring}
\end{equation}
where now $\Delta_{ij}$ are distances between jets (or subjets) in the azimuth-rapidity plane. 
The observable name originates from its geometric interpretation: radiation from colour singlets will tend to fall between the two jets, leading to values of $\mathcal{O} < 1$, while in the case of colour octets, we will tend to have $\mathcal{O} > 1$.

\subsection{$D_2$}

The variable $D_2$ \cite{Larkoski:2014gra} is defined as the ratio of two normalized $N$-point energy correlation functions (ECFs)~\cite{Larkoski:2013eya}, $e_k^\beta$:
\begin{equation}
  D_2^{(\beta)} = \frac{e_3^{(\beta)}}{(e_2^{(\beta)})^3},
\end{equation}
where $\beta$ is a parameter which we have set to $\beta = 2$. The variable is usually calculated on a large radius jet, and is useful to discriminate 2-prong jets from 1-prong jets. 
Furthermore, because of its sensitivity to soft radiation at wide angles, $D_2$ also probes colour correlations and it is therefore useful to disentangle different colour configurations.
However, we note that $D_2$ retains a correlation with the mass of the large radius jet; this may be a problem when designing a tagger free of any mass bias, see for instance~\cite{Dolen:2016kst}. We will come back to this aspect in Sect.~\ref{sec:massbias}.

\subsection{Lund jet plane}

The Lund jet plane is a theory-inspired representation of a jet~\cite{Dreyer:2018nbf}. 
It is formed parsing backwards the Cam-bridge-Aachen (C/A)~\cite{Dokshitzer:1997in,Wobisch:1998wt} clustering history of the jet. The procedure starts by undoing the final clustering step and by recording the kinematics of the splitting. The primary Lund jet plane is obtained by iterating the above procedure, always following the hardest branch in each splitting and recording the azimuth-rapidity separation of the branches involved in the splitting and the relative transverse momentum of the emission.
The Lund jet plane has been exploited in the context of vector boson~\cite{Dreyer:2018nbf}, top~\cite{Dreyer:2020brq} and Higgs~\cite{Khosa:2021cyk} tagging. Furthermore, it has been successfully measured at ATLAS~\cite{ATLAS:2020bbn} and first-principle theoretical predictions have been performed in Ref.~\cite{Lifson:2020gua}.

\section{Event simulation and selection}
\label{sec:analysis}

\begin{figure}
  \centering
  \begin{subfigure}{.35\textwidth}
    \centering
    \includegraphics[width=\textwidth]{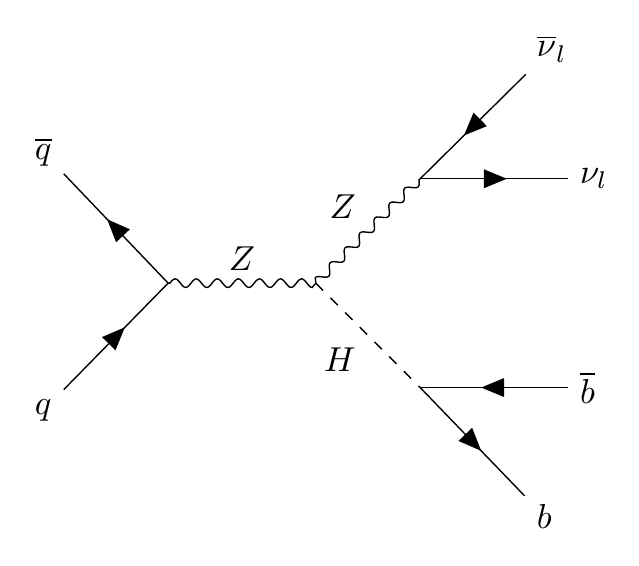}
    \caption{Signal}
    \label{fig:feysig}
  \end{subfigure}
  \begin{subfigure}{.4\textwidth}
    \centering
    \includegraphics[width=\textwidth]{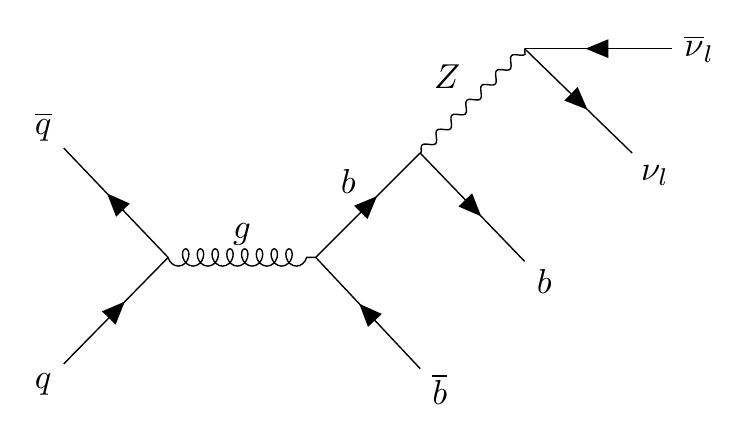}
    \caption{Background}
    \label{fig:feybkg}
  \end{subfigure}
  \caption{Representative Feynman diagrams.}
  \label{fig:feyn}
\end{figure}

We generate about 300k events for the $p p\rightarrow H(b\overline{b}) Z(\nu_\ell \overline{\nu}_\ell)$ signal and 4M events for the $ p p \rightarrow b \overline{b} \nu_\ell \overline{\nu}_\ell$ background processes, so as to have about 50k events remaining after all analysis cuts, in accordance with Table~\ref{tab:eff}, as will be detailed below. In case of signal, $b\overline{b}$ pair is produced from the decay of the Higgs boson, while it comes mainly from QCD interaction in case of background. Fig.~\ref{fig:feysig} and Fig.~\ref{fig:feybkg} show representative Feynman diagrams for the signal and the background processes, respectively.

We generate hard events using \code{MG5\_aMC@NLO v2.8.3.2}~\cite{Alwall:2014hca}, by imposing a 200 GeV cut on the $p_T$ for the neutrino pair in the final state. This is done to ensure that the events generated are firmly in the boosted regime.
These parton-level events are subsequently showered in \code{Pythia} v8.305 \cite{Sjostrand:2014zea}, including MPI and underlying events, to produce particle-level events.

Finally, rather than simulating an entire detector, \code{Delphes} v3.5.0 is used to perform a fast detector simulation \cite{Ovyn:2009tx,deFavereau:2013fsa}. This allows us to understand how the discrimination power could be affected by real-life detector effects, without having to run a computationally expensive full simulation, which in addition is strongly detector dependent. From \code{Delphes}, we extract both the Monte Carlo truth of the event, containing the particle-level information, e.g. the same one would get from a perfectly efficient detector (henceforth referred to as {\em truth}), as well as the reconstructed events including the detector effects (henceforth referred to as {\em reco}).
The \code{Delphes} simulation is run using the ATLAS card with minor modifications, described below, to fit our needs. For the truth case, we consider all visible, stable particles with $p_T > 0.5$~GeV. Instead, in the reco case, the jets are built using the simulated calorimeter towers and tracks. All electromagnetic calorimeter towers with energy $E > 0.5$ GeV and significance $S > 2.0$ and all hadronic calorimeter towers with energy $E > 1.0$ GeV and significance $S > 2.0$ are considered. Tracks are required to have $p_T > 0.5$ GeV. \code{Delphes} uses \code{FastJet} v3.3.4 \cite{Cacciari:2011ma} to perform the jet clustering.

At this point the analysis is the same in both the truth case and the reco case.
First, the constituents are clustered into jets with radius $R=1.0$  using the anti-$k_T$ algorithm \cite{Cacciari:2008gp}. For each event, we choose the jet with the highest $p_T$ as the {\em large radius jet}. We only accept the event if the large jet has $p_T > 250$ GeV and $\vert y \vert < 1.5$, because of the tracking detector acceptance.

We also cluster the constituents into smaller jets with radius $R = 0.2$. We identify those jets having $\Delta R < 0.8$ from the large jet, and call these subjets. We then proceed to identify the $b$-subjets which originate from the $b$-partons, through a process known as \emph{b-labelling}. We do this by first identifying the $b$-partons originating from the hard scattering in the event record, requiring a minimum $p_T$ of $5.0$ GeV. For each $b$-parton, we compute the distance between each $b$-parton and subjet; the subjet which is closest to the $b$-parton, provided that the distance is below 0.2, is labelled as $b$-subjet. The association between a $b$-parton and $b$-subjet is unique.

For the event to be accepted, we require two $b$-labelled subjets with $p_T > 10$ GeV. The pull variables are calculated on these two $b$-labelled subjets, and $D_2$ is calculated on the large jet. For the colour ring to be defined, there must also be a third non-$b$-subjet within $\Delta R = 0.8$ from the large jet. In a majority of cases, this third jet is not present. To avoid discarding too many events, in these cases we assign a default value of $\mathcal{O} = -1$ to the colour ring. This allows for higher statistics, but also provides useful information to the machine learning algorithms.

For the Lund jet plane, we consider large radius jet constituents and re-cluster them using C/A algorithm. Considering the declustering history of this jet, we get the primary Lund jet plane. Considering 25 $\times$ 25 pixels for each image, we put 1 or 0 in a pixel depending on if ($\ln 1/\Delta, \ln k_T$) value of the splitting falls in that pixel or not. Our implementation of the Lund jet plane is based on the one present in  \code{fastjet-contrib}~\cite{fastjetcontrib} repository.

Table~\ref{tab:eff} shows the percentage of events which pass the selections in all cases considered. The selection discards more background than signal events in both the truth and reco cases. The most important cut is the $p_T$ cut on the large radius jet, accounting for 60\% of discarded events. The second most important cut is the rapidity cut on the large radius jet which rejects 10$\%$ of the events.

\begin{table}[h]
    \centering
    \caption{The efficiencies of the analysis after cuts are applied.}
    \label{tab:eff}
    \begin{tabular}{lll}\hline 
    \hline\noalign{\smallskip}
        &  Truth & Reco  \\
        \noalign{\smallskip}\hline  \hline\noalign{\smallskip}
        Signal     & 20\% & 17\% \\
        Background & 1.6\% & 1.3\% \\
        \noalign{\smallskip}\hline \hline 
        \end{tabular}
\end{table}

\section{Discrimination performance}
\label{sec:results}

\begin{figure*}[th]
  \captionsetup[subfigure]{labelformat=empty}
  \begin{subfigure}{.33\textwidth}
    \centering
    \includegraphics[scale=0.56]{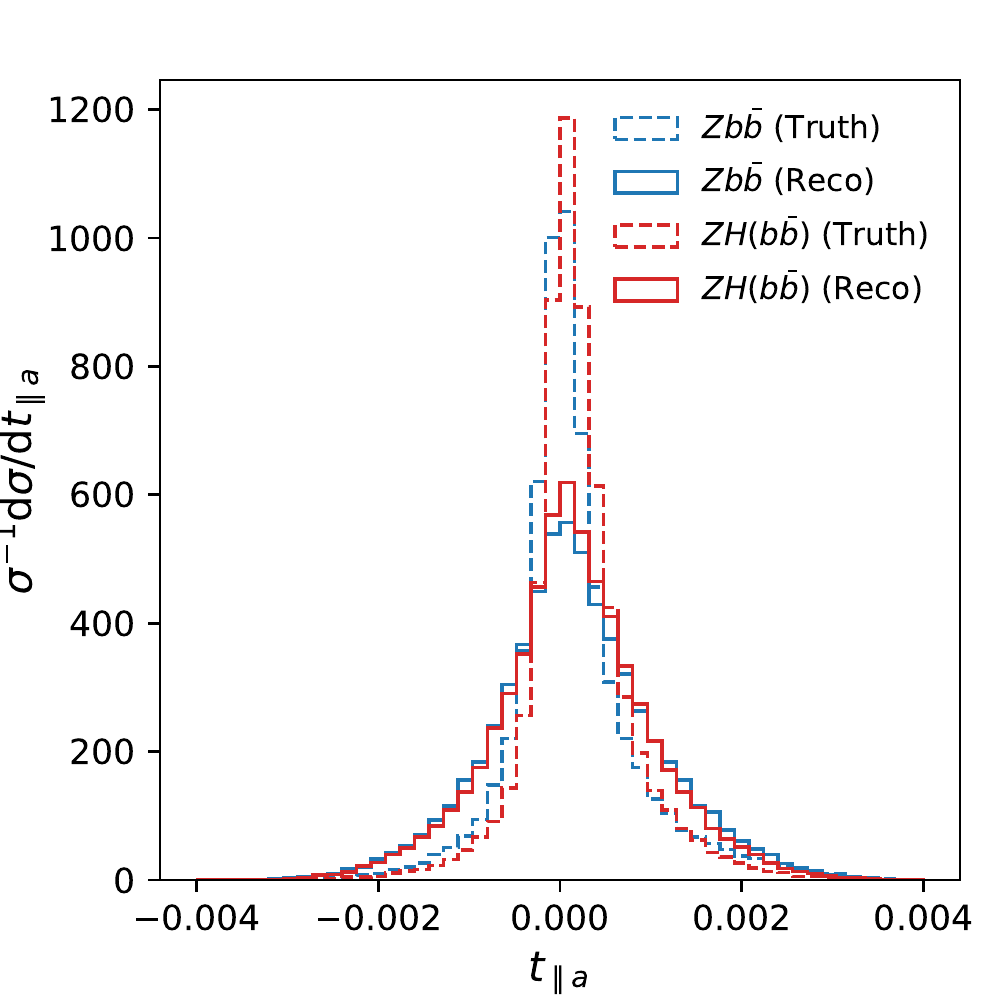}
  \end{subfigure}
  \begin{subfigure}{0.33\textwidth}
    \centering
    \includegraphics[scale=0.56]{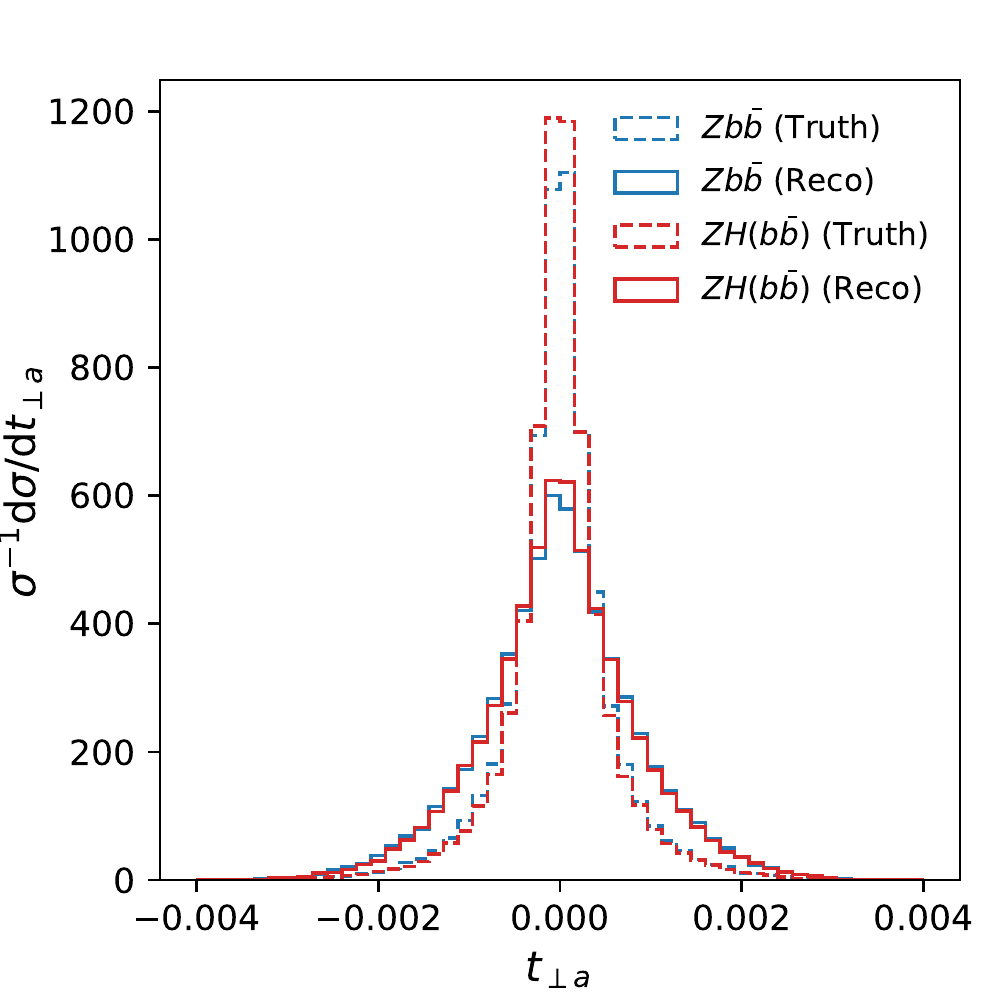}
  \end{subfigure}
  \begin{subfigure}{.33\textwidth}
    \centering
    \includegraphics[scale=0.56]{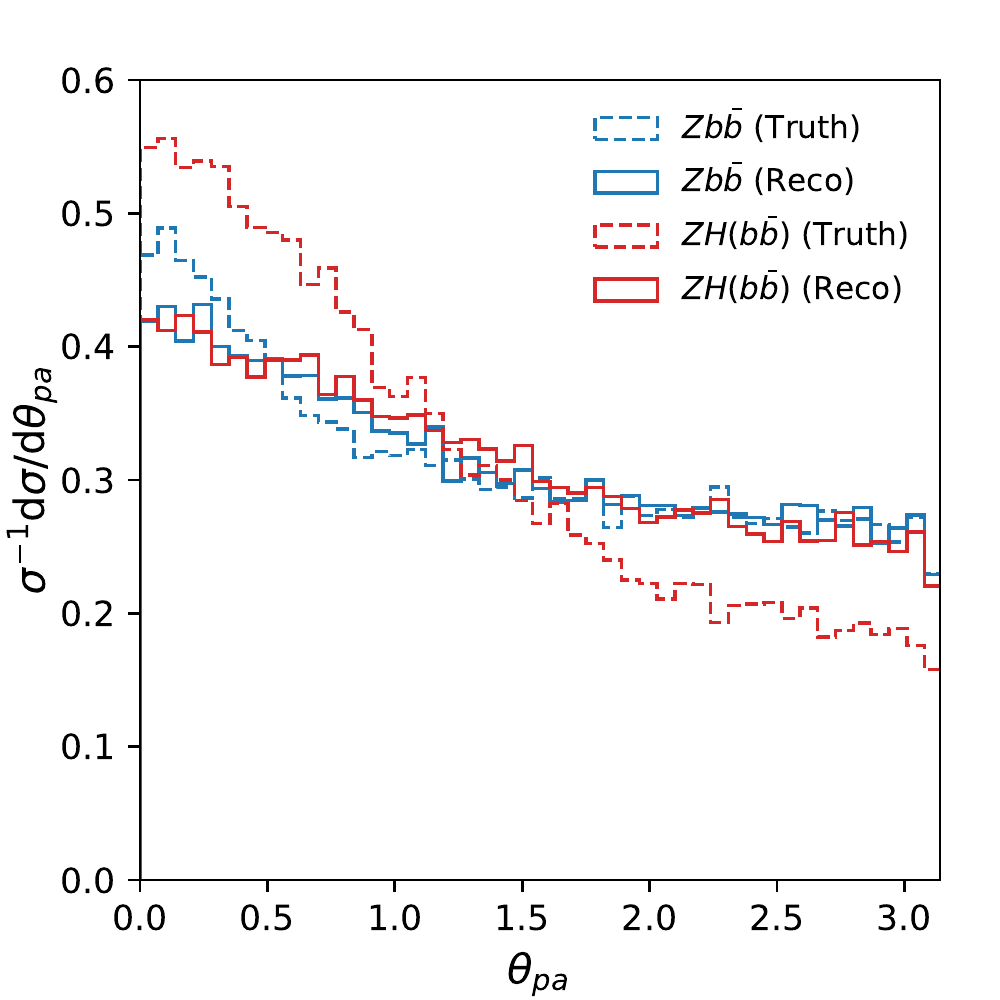}
  \end{subfigure}
  \begin{subfigure}{.33\textwidth}
    \centering
    \includegraphics[scale=0.56]{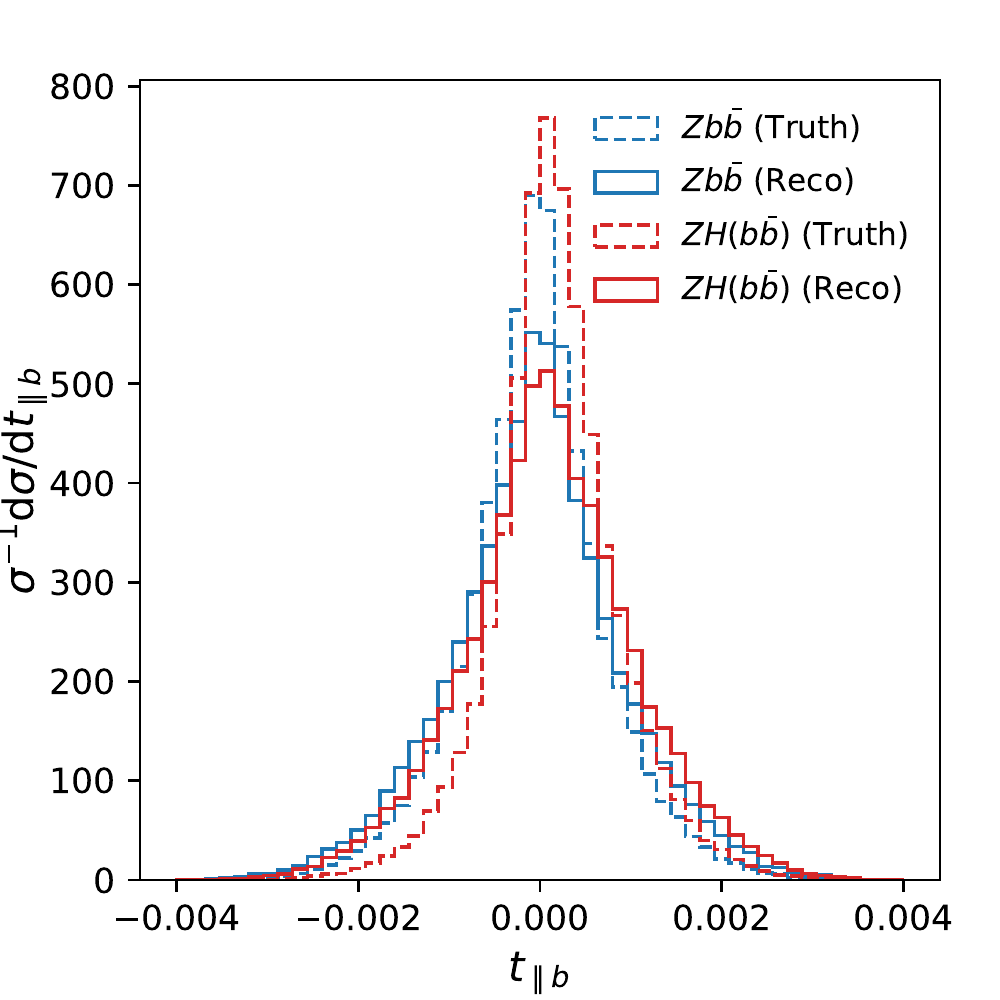}
  \end{subfigure}
  \begin{subfigure}{0.33\textwidth}
    \centering
    \includegraphics[scale=0.56]{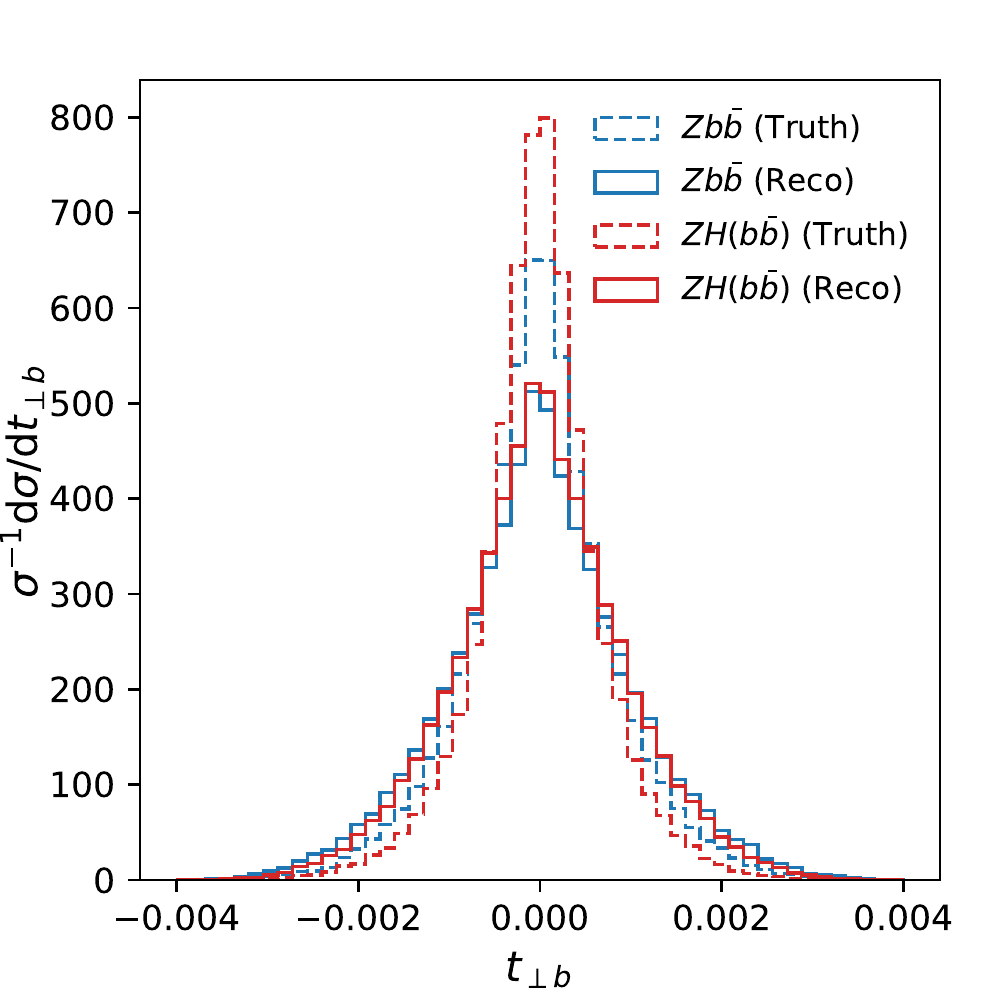}
  \end{subfigure}
  \begin{subfigure}{.33\textwidth}
    \centering
    \includegraphics[scale=0.56]{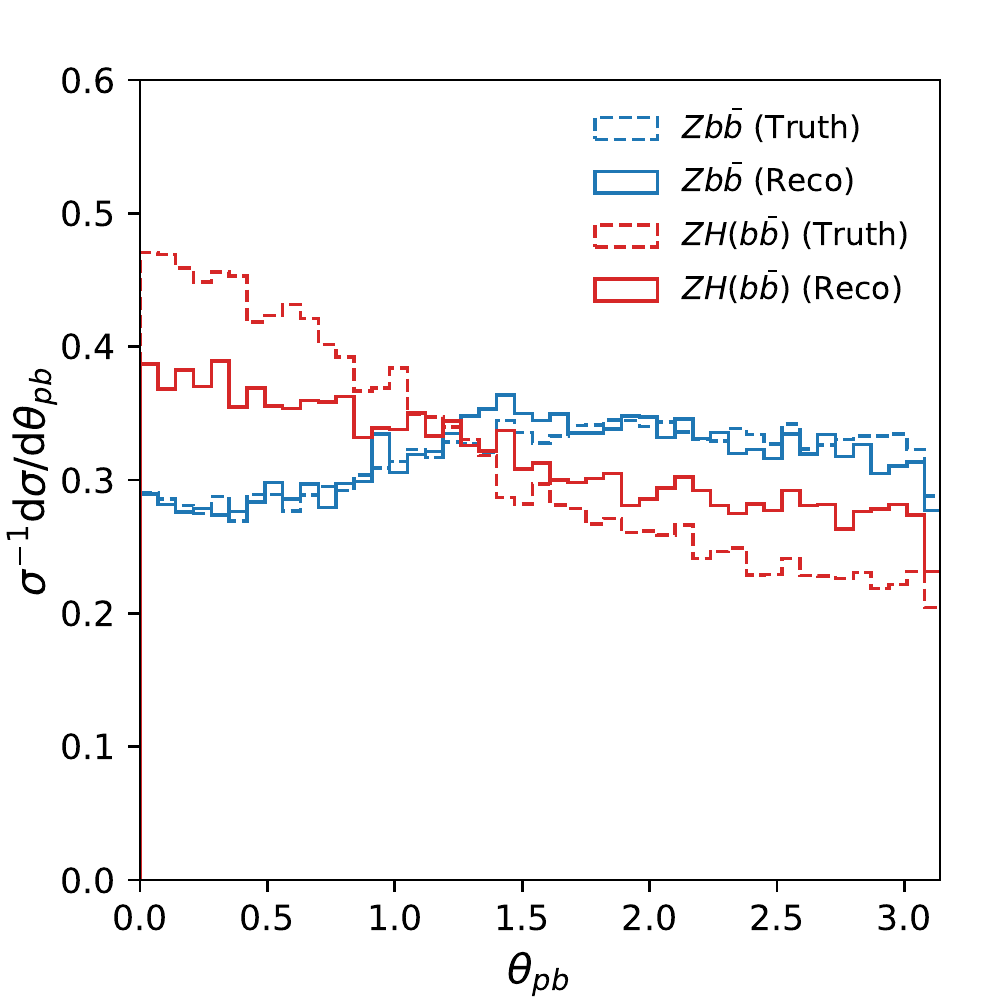}
  \end{subfigure}
  \begin{subfigure}{0.33\textwidth}
    \centering
    \includegraphics[scale=0.56]{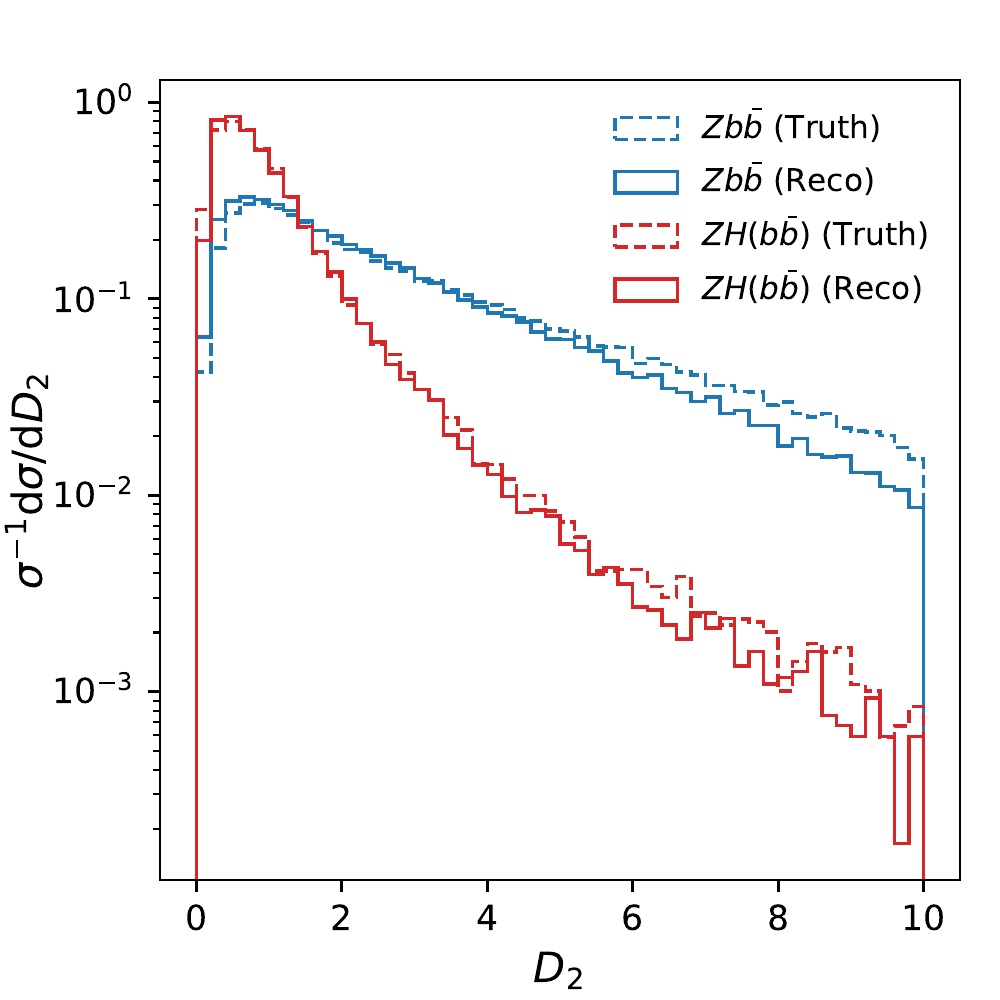}
  \end{subfigure}
  \begin{subfigure}{.33\textwidth}
    \centering
    \includegraphics[scale=0.56]{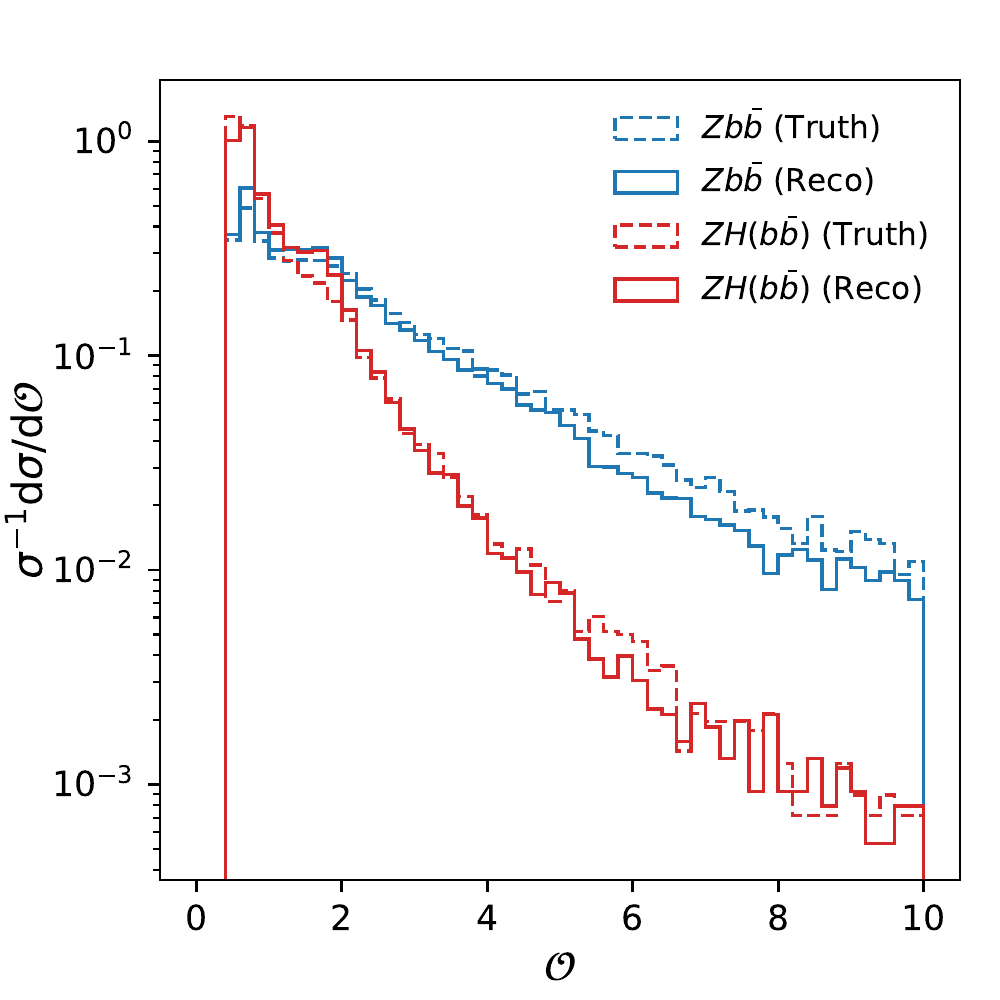}
  \end{subfigure}
  \begin{subfigure}{.33\textwidth}
    \centering
    \includegraphics[scale=0.56]{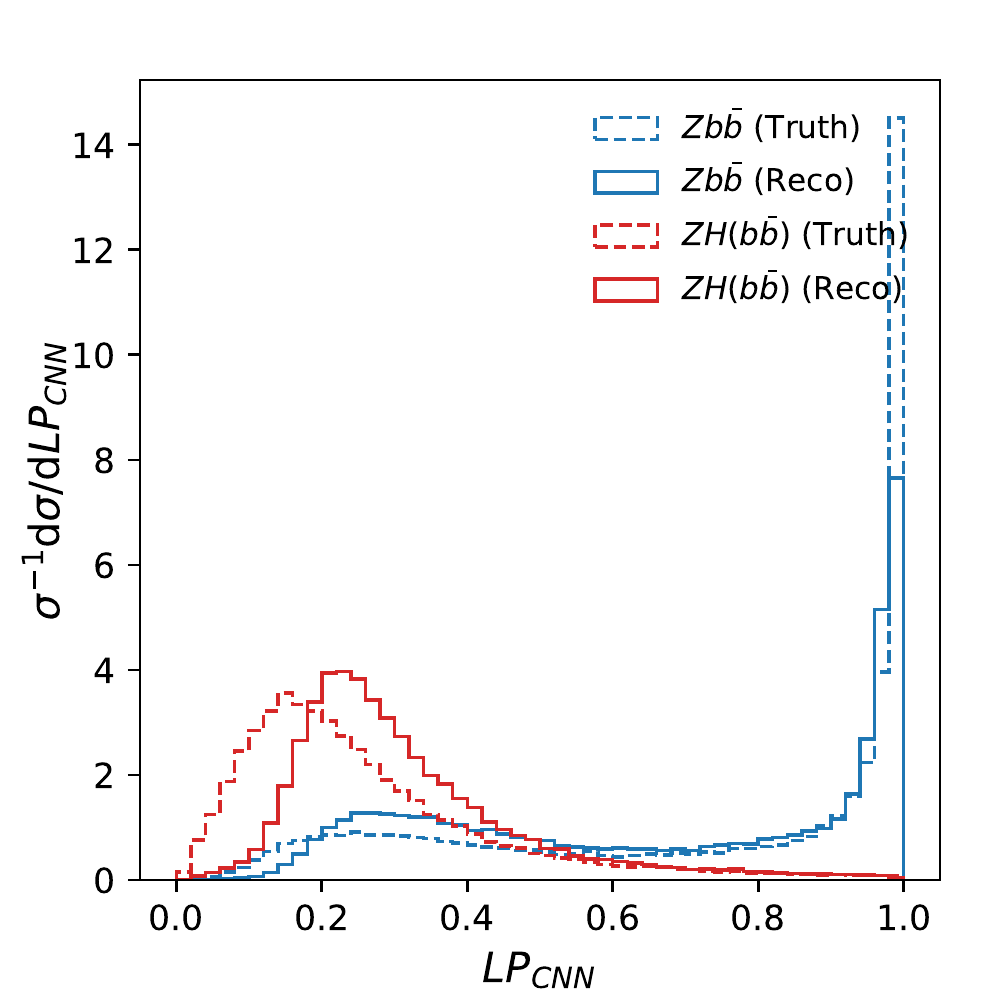}
  \end{subfigure}
  \caption{The distributions of the colour sensitive variables and Lund jet plane after the selection cuts. Signal ($ZH(b\bar b)$) and  background ($Z b \bar b$) distributions are shown in red and blue, respectively, for both truth and reco cases.}\label{fig:sgn_bkg_var}
\end{figure*}

After event selection, we are ready to evaluate observables on the selected events.
We first show in Fig.~\ref{fig:sgn_bkg_var} the normalised distributions for the eight colour sensitive (CS) observables introduced in Sect.~\ref{sec:colobs}, both for signal and background, and at the truth and reco level.
By just looking at the plots, we can appreciate the strong discrimination power of the $\mathcal{O}$ and $D_2$, which is retained at the reco level. 
Instead, the observables related to the jet pull vector are more affected by detector effects: there is a visible difference in $t_{\parallel i}$, $t_{\perp i}$ and $\theta_{pi}$, both for the leading jet $a$ and the sub-leading jet $b$, between the truth and the reco cases. 
In particular, the pull angle observables $\theta_{pa}$ and $\theta_{pb}$ seem to be good discriminants at the truth level, but detector effects noticeably flatten the signal distribution, hence leading to a worsening of the discrimination power. 

\begin{figure*}[th]
  \centering
  \includegraphics[scale=0.6]{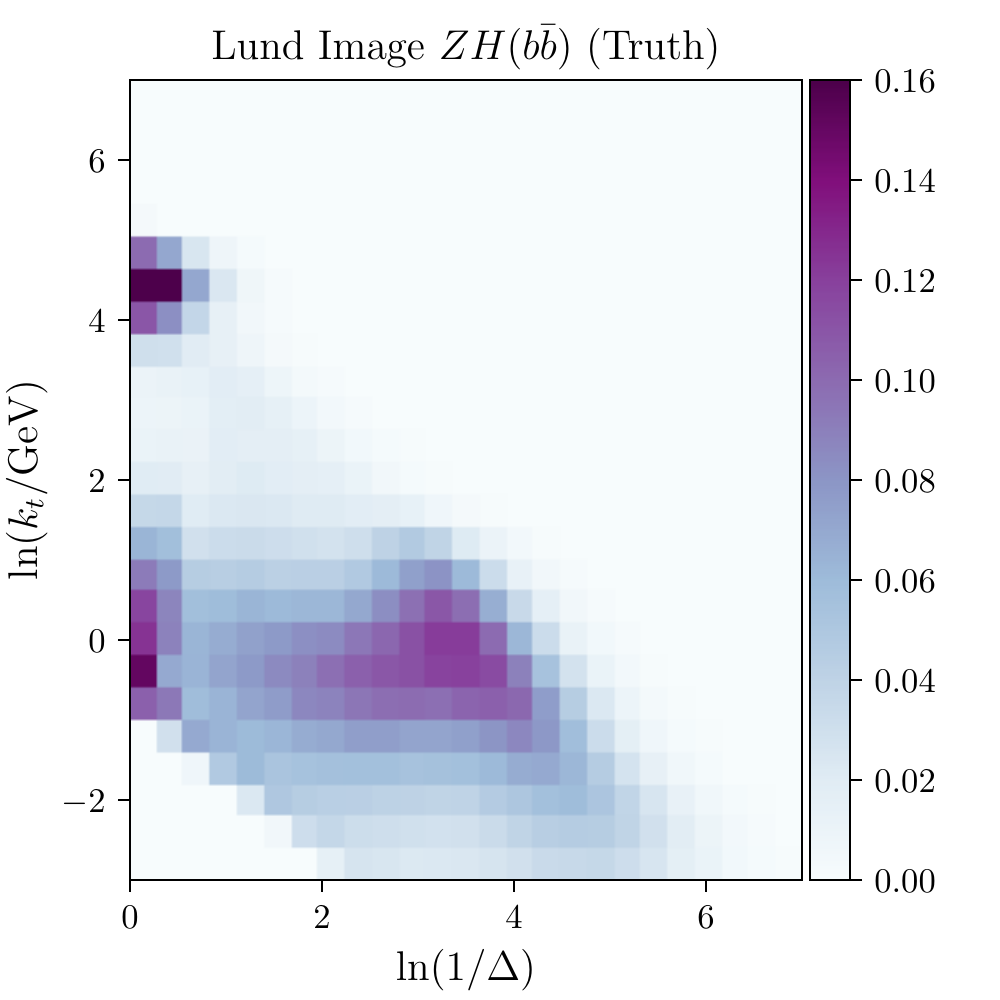}
   \includegraphics[scale=0.6]{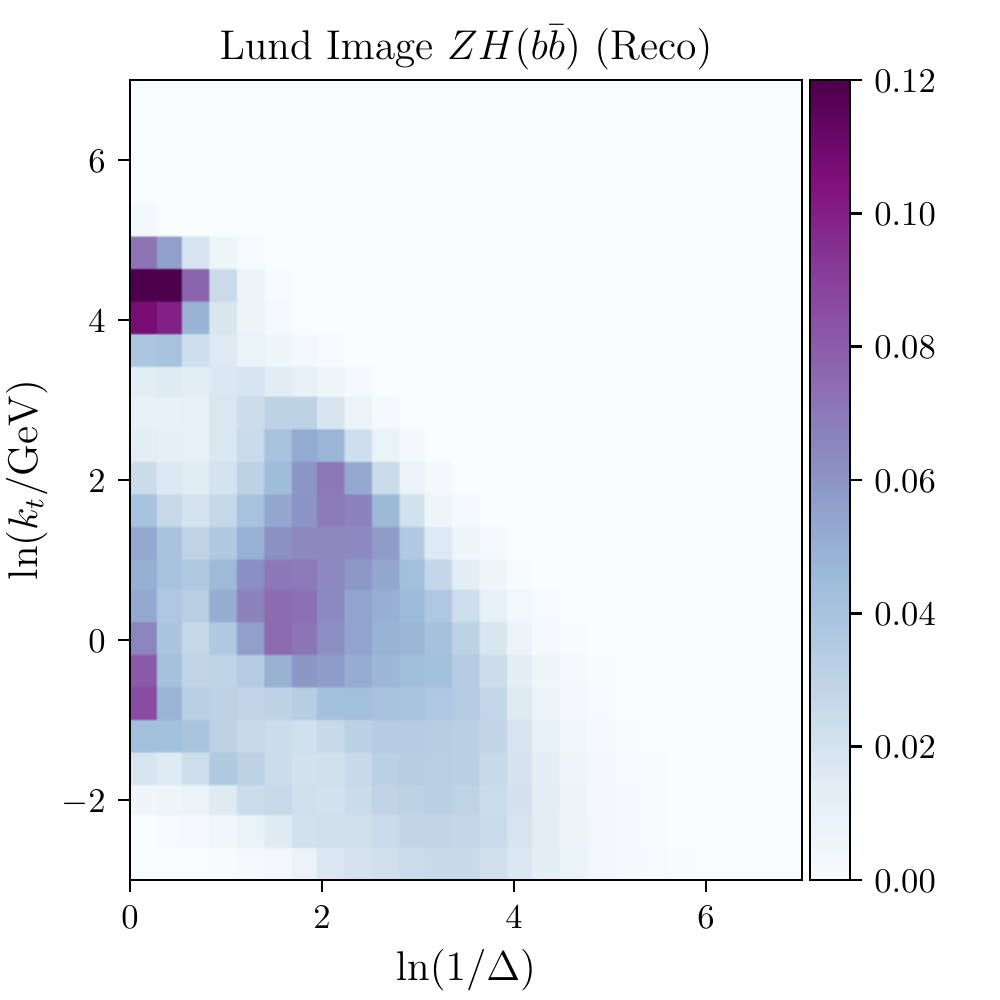}\\
  \includegraphics[scale=0.6]{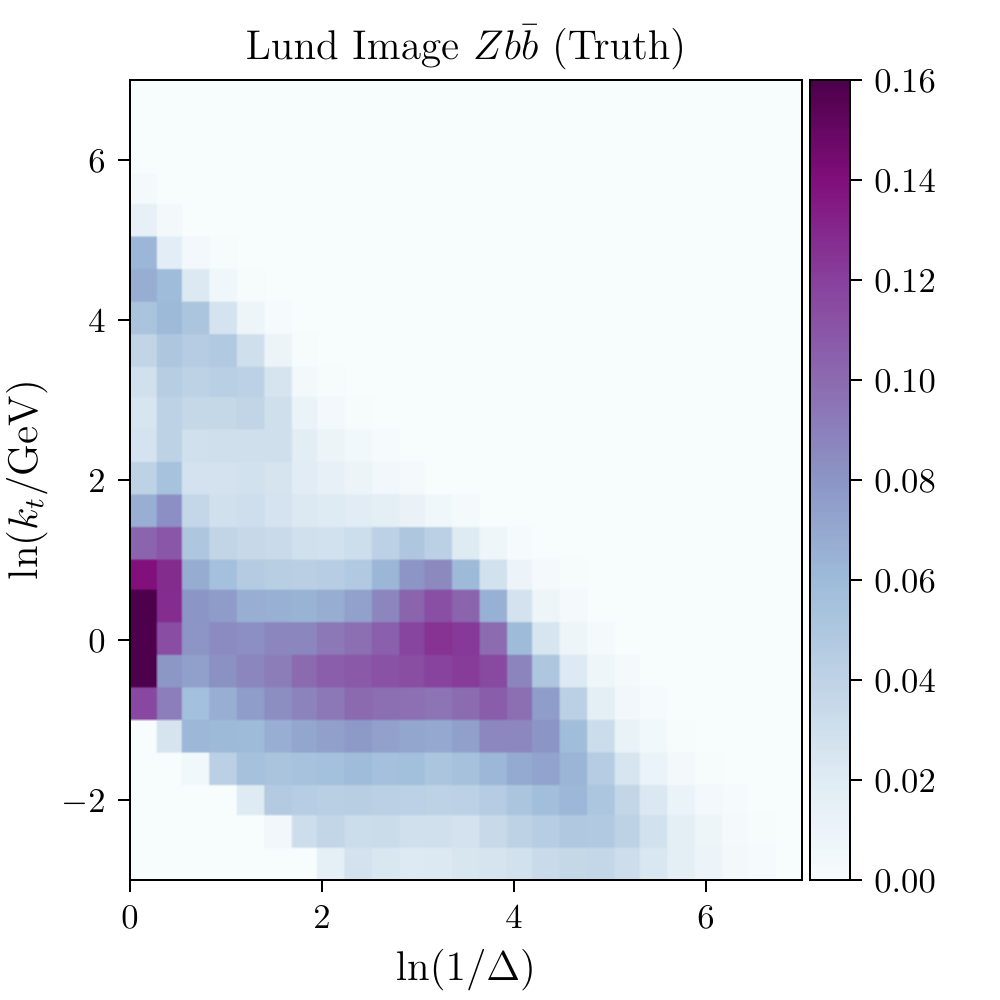}
   \includegraphics[scale=0.6]{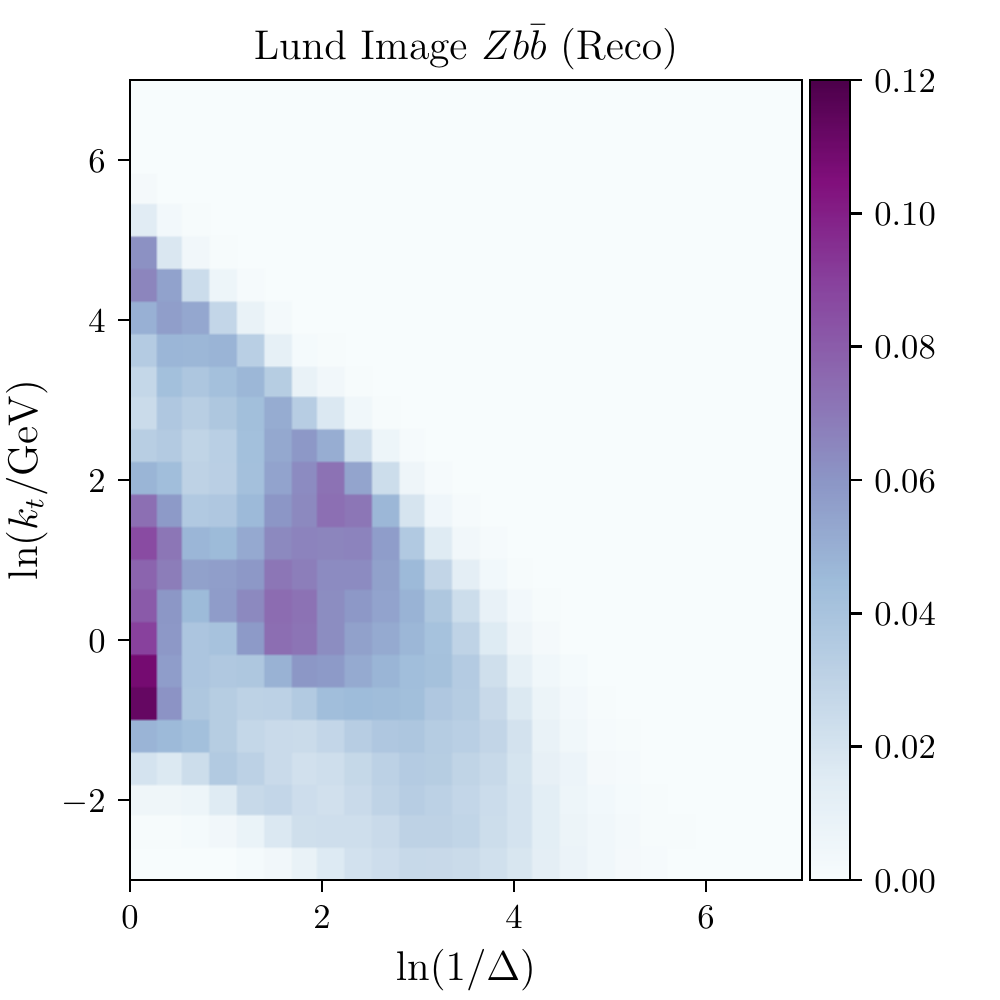}
  \caption{Averaged primary Lund jet plane images for $ZH(b\bar{b})$ and $Zb\bar{b}$ in the truth and reco case.}
  \label{fig:lundimages}     
\end{figure*}

We then show in Fig.~\ref{fig:lundimages} the averaged Lund images for the signal and background process, in the truth and reco cases respectively. We note that detector effects lead to an overall decrease of the image quality, in the sense that the distinctive features of the truth case are still present, but in the reco case there is additional radiation for middle values of $\Delta$ and $k_t$ for both the signal and the background events. However, the high density patch at large $\Delta$ and high $k_t$ in the case of the signal is still clearly visible by eye also in the reco case.

After having determined the distributions of the CS observables and the Lund jet images, we now use these as inputs to ML algorithms in order to build combined classifiers. Specifically, we train a BDT\footnote{We have also tried a neural network, which returns similar results in all the cases analysed. Hence, we only report results obtained with the BDT.} on the CS observables, whereas Lund images are classified using a CNN. More details about these methods and architectures are provided in~\ref{sec:appNN}. The output distribution of the CNN Lund jet plane classifier (\LP) is shown in Fig.~\ref{fig:sgn_bkg_var}. In the following, we consider also the combination of (some of) the CS and \LP observables, in order to improve the total discrimination power; in such cases, we adopt a two-step procedure, by using the output of the CNN Lund jet plane classifier as an additional input to the BDT. 

\begin{figure*}[h]
  \centering
    \centering
     \includegraphics[width=0.45\textwidth]{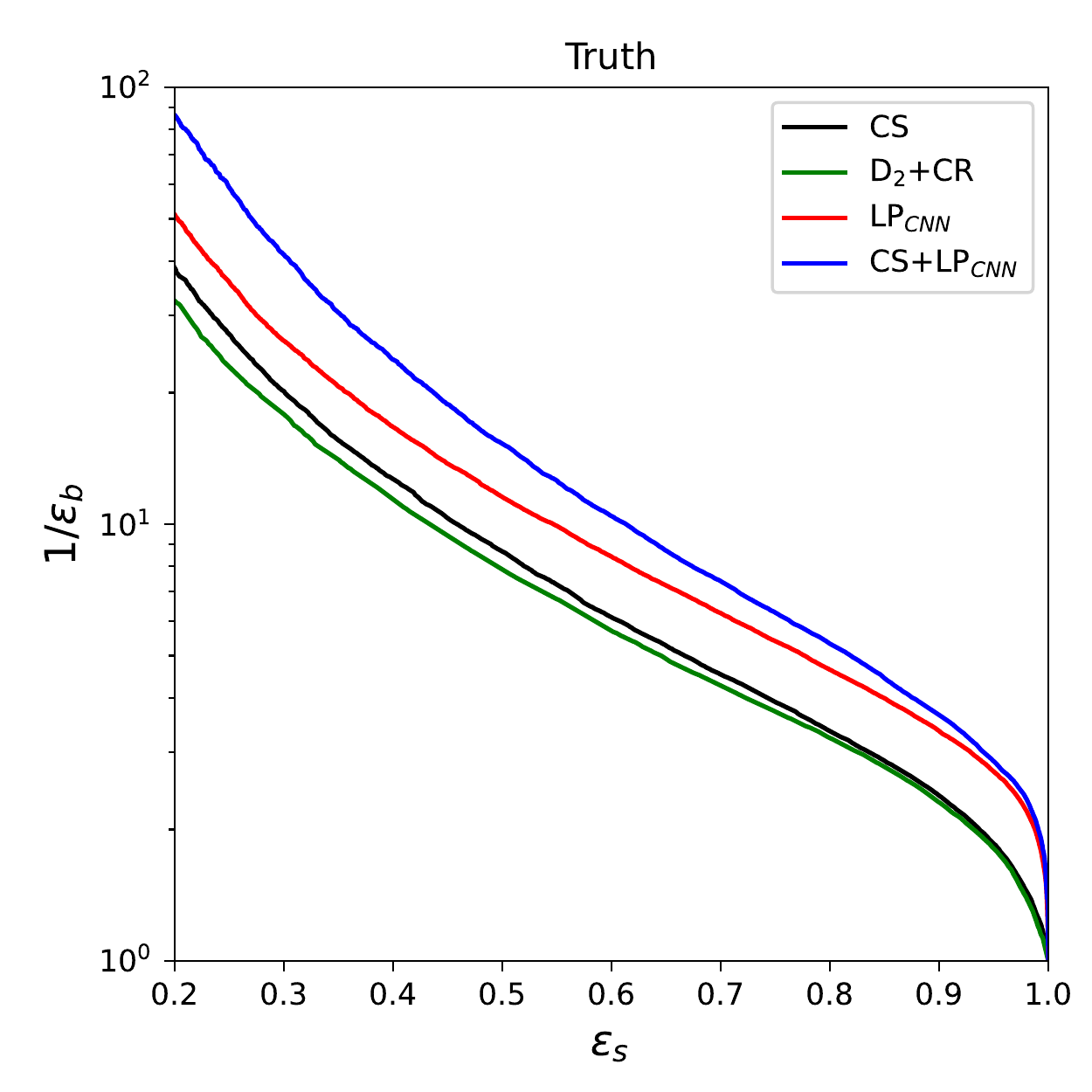}
     \includegraphics[width=0.45\textwidth]{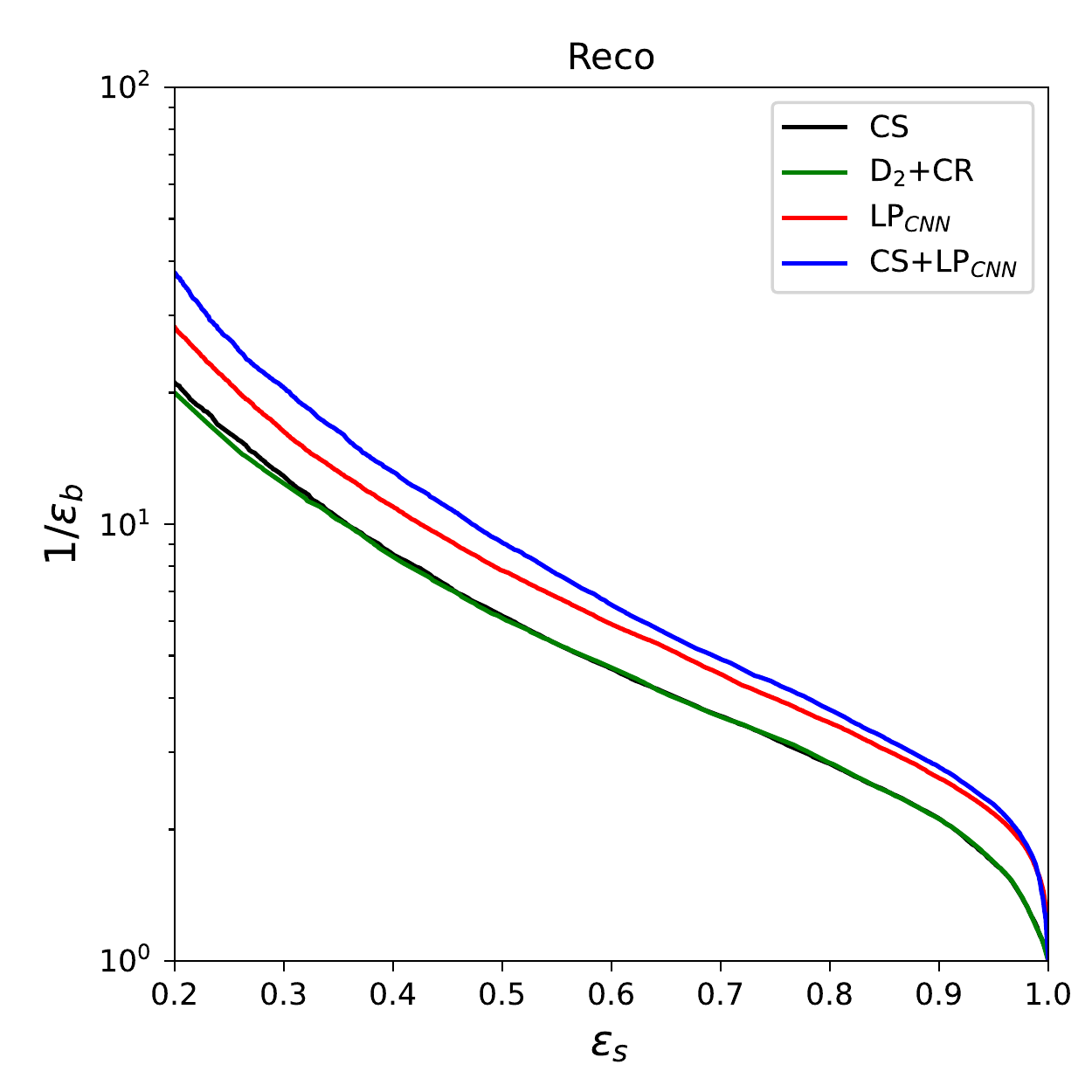}
   \caption{The ROC curves showing background rejection as a function of signal efficiency for the truth (left) and reco case (right) for CS variables, \LP and the combined cases.}
  \label{fig:ROC}
\end{figure*}

In Fig.~\ref{fig:ROC}, we show the receiver operating characteristic (ROC) curves for several combinations of observables. Namely, we consider all the colour sensitive observables (CS) or just the $D_{2}$ and the colour ring ($D_{2}$+CR), combined through a BDT; the \LP; the combination of all the CS observables with the \LP (CS+\LP), by means of the two-step procedure explained above.
For each curve in Fig.~\ref{fig:ROC}, we report the value of the area under the ROC curve (AUC) in Table~\ref{tab:AUC}, both for the truth and the reco cases. A perfect classifier would have $\text{AUC}=1$, whereas a random classifier is associated with $\text{AUC}=1/2$.

\begin{table}[h!]
  \caption{Area under the ROC curves for different combination of observables.}
  \label{tab:AUC}
  \centering
  \begin{tabular}{ccc}
    \hline \hline 
    \* & \multicolumn{2}{c}{AUC - Test Sample} \\ 
    \* & Truth & Reco \\ 
    \hline \hline 
    CS observables & 0.826 & 0.788 \\ 
    $D_{2}$+CR & 0.817 & 0.787 \\
    \LP & 0.876 & 0.828 \\  
    CS + \LP & 0.893 & 0.846 \\
    \hline  \hline 
  \end{tabular} 
\end{table} 

As expected, we first observe a worsening of the discrimination power by moving from the truth case to the reco case, as can also be seen by comparing the value in the left and in the right column of Table~\ref{tab:AUC}. However, the performance after taking into account detector effects is still good for most of the combinations, close to 0.85 for the CS+\LP combination.
Furthermore, we see that most of the discriminating power of the set of CS observables is actually coming from the combination of $D_2$+CR alone. This is in agreement with what observed at the level of distributions at the beginning of Sect~\ref{sec:results}: the jet pull observables, including the pull angle, seem not to add any additional information useful for classification. At reco level, their discrimination power is almost unnoticeable.
Moving to the combinations involving the Lund jet plane, we observe that the Lund jet plane alone is performing better than the whole set of CS observables, especially in the region of high signal efficiencies.
When we combine \LP with the CS observables, we see a noticeable improvement of the overall classification power, with a value of AUC equal to 0.893 in the truth case and 0.846 in the reco case.

Finally, in Table~\ref{tab:rank} we rank the variables based on their importance in the BDT, both in the truth and reco case. The ranking presented here also includes the output of the Lund jet plane CNN as an additional input. \LP is the most discriminating variable, both in the truth and in the reco case.
It is followed in order by $D_2$ and the colour ring $\mathcal{O}$. The jet pull variables are all of similar importance, at the bottom of the ranking score. For the reconstructed case, $\mathcal{O}$ gains additional importance with respect to the pull variables.

We end this Section with a comment about the usage of the pull angle. As already mentioned in Sect.~\ref{sec:colobs}, the pull angle $\theta_{p}$ is only Sudakov safe.
One may wonder what is the effect of keeping only its (IRC-safe) projections $t_{\parallel}$ and $t_{\perp}$, instead of using all the three observables $t_{\parallel}$, $t_{\perp}$ and $\theta_{p}$ as input in the BDT, as we have done.
Unsurprisingly, by looking at the correlation matrix in the BDT, we observe a strong correlation between these variables. Given that the variables derived from the jet pull vector(s) have a small influence on the overall performance, a variant of the tagger could be conceived, with comparable performance, by dropping $\theta_{p}$ among the list of inputs to the BDT.

\begin{table}[th]
    \caption{BDT observable ranking for the truth and reco cases.}
    \label{tab:rank}
    \centering
    \begin{tabular}{ccccc}
      \hline \hline
      \multicolumn{5}{c}{Observable Ranking} \\ 
      \hline
     & \multicolumn{2}{c}{Truth} & \multicolumn{2}{c}{Reco} \\ 
      \hline 
      Rank & Obs. & Importance  & Obs. & Importance\\ 
      \hline 
      1 & \LP  & 6.6 $\times 10^{-1}$& \LP & 4.8$\times 10^{-1}$ \\
      2 & $D_2$ & 1.4 $\times 10^{-1}$ & $\mathcal{O}$ & 1.0$\times 10^{-1}$  \\ 
      3 & $\mathcal{O}$ & 5.7$\times 10^{-2}$  & $D_2$ & 9.3$\times 10^{-2}$  \\ 
      4 & $\theta_{pb}$ & 3.0$\times 10^{-2}$ &  $\theta_{pb}$ & 7.0$\times 10^{-2}$\\ 
      
      5 & $\theta_{pa}$  & 2.9$\times 10^{-2}$ & $\theta_{pa}$ &  6.5$\times 10^{-2}$\\ 
      6 & $t_{\parallel b}$ & 2.6$\times 10^{-2}$ & $t_{\perp b}$ & 
      6.0$\times 10^{-2}$  \\ 
      7 & $t_{\parallel a}$ & 2.4$\times 10^{-2}$ & $t_{\parallel a}$  & 4.5$\times 10^{-2}$\\ 
      8 & $t_{\perp b}$ & 1.9$\times 10^{-2}$ &$t_{\perp a}$ & 4.3$\times 10^{-2}$ \\ 
      9 & $t_{\perp a}$ & 1.0$\times 10^{-3}$  & $t_{\parallel a}$  & 3.3$\times 10^{-2}$ \\ 
      \hline \hline
    \end{tabular} 
    \end{table}

\section{Invariant mass dependence}
\label{sec:massbias}

Since our goal is to develop a tagger purely sensitive to the colour configuration of the decaying particle, in order to be applied to other contexts (such as $Z$ or $W$ boson hadronic decays), ideally our procedure should be insensitive to the invariant mass of decaying system, specifically to the invariant mass of the pair of $b$-jets.

\begin{figure*}[h]
  \begin{subfigure}{0.33\textwidth}
    \centering
    \includegraphics[width=\textwidth]{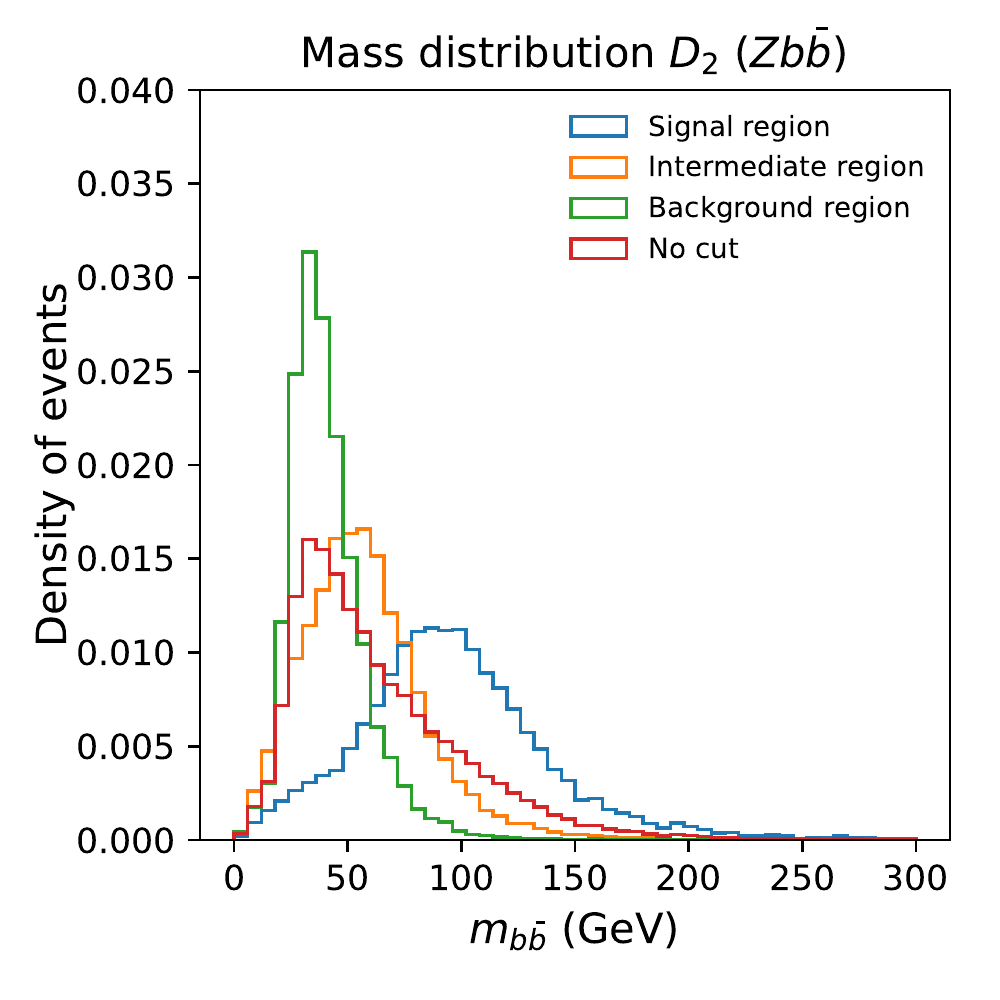}
    \caption{}
    \label{mass d2}
  \end{subfigure}
  \begin{subfigure}{0.33\textwidth}
    \centering
    \includegraphics[width=\textwidth]{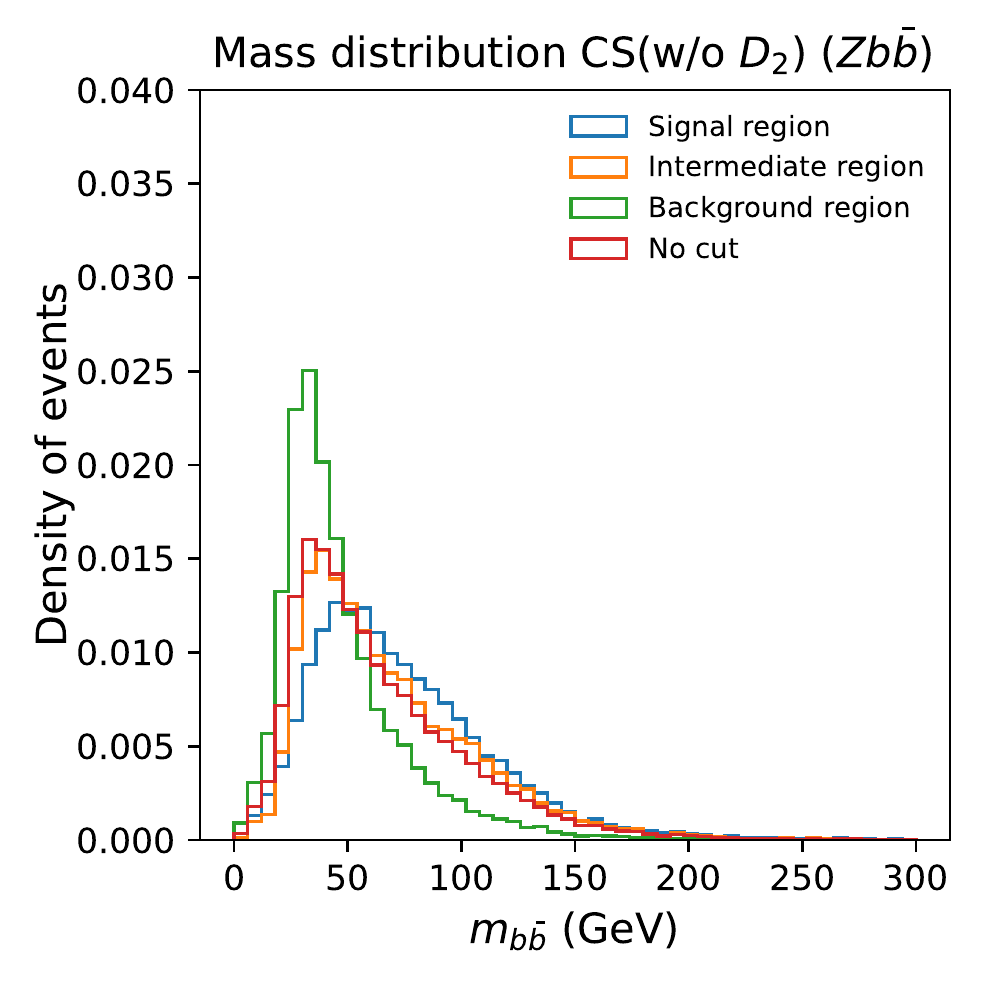}
    \caption{}
    \label{bdt w/o d2}
  \end{subfigure}
  \begin{subfigure}{0.33\textwidth}
    \centering
    \includegraphics[width=\textwidth]{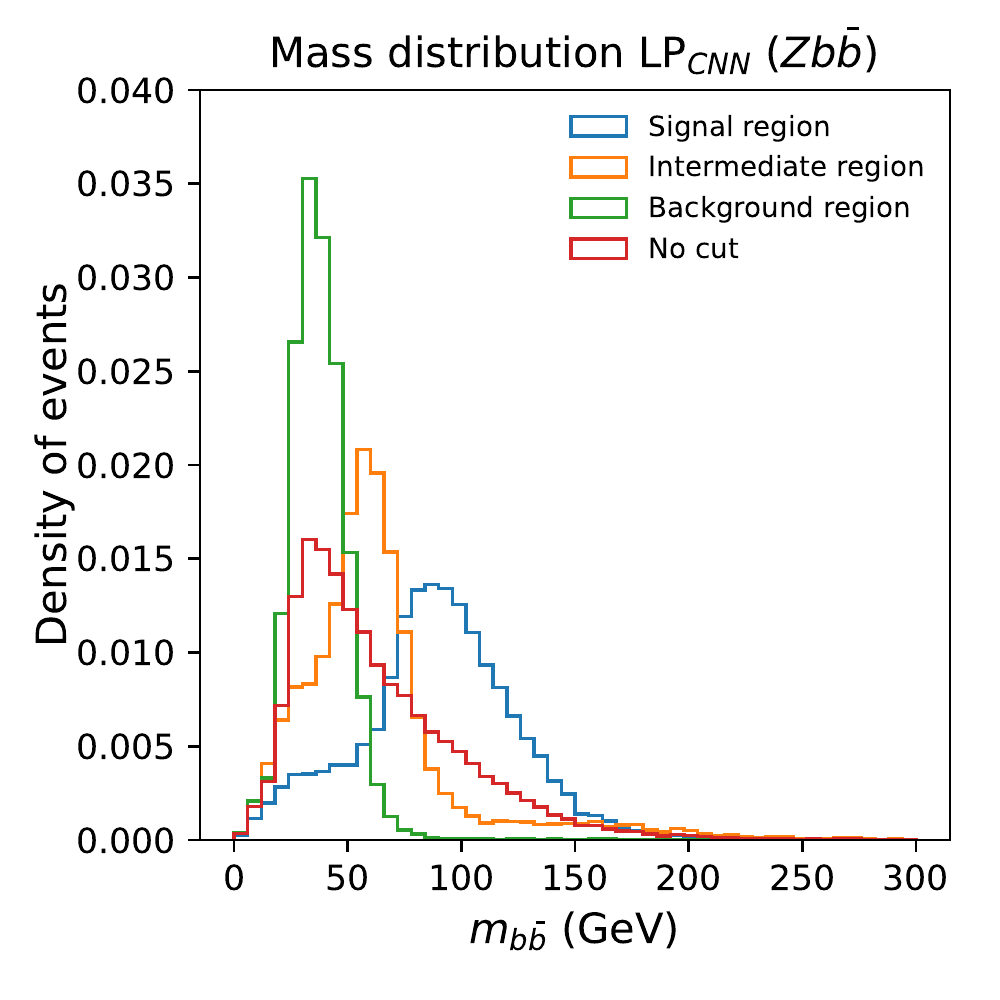}
    \caption{}
    \label{lund_mass}
  \end{subfigure}
  \caption{The distribution of the invariant mass of the $b$-jets pair in the reco background sample for (a) different cuts on $D_2$, (b) different ranges of the BDT output for the CS without the inclusion of $D_2$, (c) and for different cuts on Lund plane CNN output.}
  \label{fig:bdtmass}
\end{figure*}

In Fig.~\ref{fig:bdtmass} we show the distribution of the invariant mass measured on the whole set of background events and on three subset of events, each of the same size, corresponding to signal-enriched, intermediate and background-enriched regions.
These regions are defined by means of a set of cuts on different discriminant variables: $D_2$ alone (Fig.~\ref{mass d2}), combined BDT with CS but $D_2$ (Fig.~\ref{bdt w/o d2}), Lund plane CNN (Fig.~\ref{lund_mass}).
We show results for the reco case only, since the ones in the truth case are similar.
In an ideal scenario, a cut on the discriminating variable should not also concomitantly imply a cut on the invariant mass of the system, hence the curves for the three regions should overlap, and agree with the curve without any cut.

We find that the sensitivity to the invariant mass is introduced mainly through the $D_2$ observable, which is highly correlated to the value on the mass, as it is clear from Fig.~\ref{mass d2}. Such a correlation has been already investigated in the literature~\cite{Dolen:2016kst, Moult:2017okx}.
By removing $D_2$ from the CS input variables of the BDT, the mass bias is greatly reduced, as can be observed by comparing Fig.~\ref{mass d2} and Fig.~\ref{bdt w/o d2}.
However, given the fact that the $D_2$ observable is ranked as one of the most important (see Table~\ref{tab:rank}), the removal of this variable comes at the price of loosing a good part of efficiency.

Finally, it is interesting to study whether the \LP alone retains or not a dependence on the invariant mass of the decaying system. This is shown in Fig.~\ref{lund_mass}. 
Unfortunately, the output of the \LP appears to be notably correlated to the invariant mass of the pair of $b$-jets. In order to better understand this behaviour, by looking at the correlation matrix of the BDT with the CS variables and the \LP as input, we note that \LP and $D_2$ are largely correlated. 
Hence, the behaviour we observe in Fig.~\ref{lund_mass} for \LP can be related to the known behaviour of $D_2$ of Fig.~\ref{mass d2}.
Given the fact that \LP is our best discriminating variable, considering removing it from the combination comes at the price of loosing a good part of discrimination power.

\section{Conclusions}
\label{sec:concl}

In this paper, we have investigated the problem of distinguishing the $b$-jets originating from the decay of a colour singlet from those originating from the pure QCD background (mostly through $g \to b\bar{b}$ collinear splitting).
We have focused on the signal process $p p \rightarrow H(b\overline{b}) Z(\nu_\ell \overline{\nu}_\ell)$, but we are confident that our strategy is valid in a more general context. 
Specifically, we have trained a BDT architecture on eight high-level, colour sensitive observables, in order to develop a combined colour tagger. We have also explored the discrimination performance of a CNN architecture trained on Lund jet plane images. Finally, we have combined the high-level observables and the output of the Lund jet plane CNN in a common BDT architecture.
We have also performed a fast detector simulation in order to better assess the experimental feasibility of this tagging strategy. Namely, we have compared individual distributions and the final performance of the tagger before and after the inclusion of detector effects.

We have found a good discrimination power for our combination of colour sensitive observables with the output of the Lund jet plane CNN (AUC~=~0.893), slightly deteriorated when including detector effects (AUC~=~0.846). The Lund jet plane alone has been proven to be a powerful $Hbb$ tagger even in presence of detector effect, thus extending the results of Ref.~\cite{Khosa:2021cyk}, and when combined with theoretically motivated single-variable observables, such as $D_2$ or the colour ring, the overall performance appreciably improves.
In the end, we have shown that our tagger, which is a combination of several theory-driven single-variable observables with a representation of radiation pattern within a jet, is not only effective in theory, but also shows promising prospects for application to experimental analyses.  

We have also studied to what extent the tagger is sensitive to the mass of the decaying particle. In the case of the colour sensitive observables, the mass bias has been shown to be ultimately due to the $D_2$ input variable, as already known from the literature.
However, we have also found that the Lund plane CNN retains a large mass bias, and to the best of our knowledge this has not been pointed out in the literature so far.
The elimination of these variables come at the cost of classification efficiency, especially in the case of the Lund plane CNN. Further studies are needed in order to understand how to remove such a mass bias.
For instance, one could plan to explore  techniques similar to the ones presented in Refs.~\cite{Dolen:2016kst, Moult:2017okx}, based on a debiasing {\em a posteriori}.

Even if a fast detector simulation offers a good starting point in order to assess the feasibility of a tagging strategy, in the end a full detector simulation within a more defined experimental context would be required.
Such a more realistic scenario would also entail the inclusion of the efficiency of real-life existing $b$-tagging algorithms used by the experimental collaborations, whereas in this paper we have assumed a $b$-tagging algorithm with 100\% efficiency.
Further, although the main scope of this study is to assess the performance of relatively simple observables and ML architectures for $Hbb$ tagging, it is interesting to investigate the importance of physics information beyond the primary Lund plane, which can be exploited by including planes originated by secondary splittings. In this context, we have made a preliminary investigation using the LundNet-5 model of Ref.~\cite{Dreyer:2020brq}, which employs graph neural networks, and obtaining an improved performance corresponding to AUC~=~0.925 and AUC~=~0.894, for truth and reco, respectively.  

We leave the implementation of these suggestions, as well as the opportunity to consider new high-level observables as input for our tagger, for future studies. 
\begin{acknowledgements}
The work of CK and SM is supported by Universit\`a di Genova under the curiosity-driven grant ``Using jets to challenge the Standard Model of particle physics'' and by the Italian Ministry of Research (MUR) under grant PRIN 20172LNEEZ and under the project ``Dipartimenti di Eccellenza''. AR acknowledges support from DESY (Hamburg, Germany), a member of the Helmholtz Association HGF and from the University of Z\"{u}rich. 
\end{acknowledgements}

\appendix
\section{Details about ML algorithms}
\label{sec:appNN}

In this Appendix, we report some details about BDT parameters and CNN architecture adopted in the analysis.

\subsection{BDT parameters}
The Boosted Decision Tree (BDT) has been implemented using the \code{ROOT} Toolkit for Multivariate Analysis (TMVA) library~\cite{physics/0703039}.
The BDT is made up of 50 trees, with a maximum depth of 5 and minimum node size set to 2.5\% of the total number of events. The Gini index is used as the optimisation criterion. AdaBoost has been chosen as the boosting model and the number of cuts is set to 80. We use a 50/50 train/test sample, and there is no downsampling.

\begin{table}\label{sec:6:BDT}
\centering
\begin{tabular}{cc}
  \hline \hline
  Parameters & Value \\ 
  \hline \hline 
  No. of Trees & 100 \\ 
  Max Depth & 3 \\ 
  MinNodeSize & 2.5\% \\ 
  Boost Type & AdaBoost \\ 
  Train/Test & 50/50 \\
  No. of Cuts & 200 \\  
  Downsampling & No \\ 
  \hline \hline
\end{tabular}
\caption{BDT parameters used for the truth and reco data set.}
\end{table}

\subsection{CNN architecture}
 We used CNN (implemented using Keras~\cite{chollet2015keras}) for the Lund jet images data set. Balanced data set for the  binary classification is used; $70\%$ for the training, $15\%$ for the validation and $15\%$ for the testing. We tried several models with different combinations of hyper-parameters and the best architecture is described in the Table~\ref{CNN-architecture}. Here $N_i, i=1\dots 4$ denotes the number of filters in the corresponding convolutional layer. Filter size is $3 \times 3$ in all the convolutional layers. Activation function `relu' and `softmax' are used for the intermediate and last layer, respectively. For the optimisation, `adam'  optimiser is used. We used pooling (MaxPooling)
 operation after the second and fourth convolutional layer for image downsizing. 
 Note that CNN architecture used is same for truth and reco data except the dropouts. The numbers mentioned in the brackets are the dropouts strength used for the reco data set. 

\begin{table}
\centering
\begin{tabular}{lccccc}
\hline
\hline Parameter & Value  \\
\hline
\hline
$N_1$ Conv2D   & 30              \\
$N_2$ Conv2D   & 30      \\
Dropouts       & -- (0.3)       \\
$N_3$  Conv2D  &  30               \\
Dropouts       &-  (0.3)       \\
$N_4$  Conv2D  & 10         \\
Dropouts       &-  (0.1)        \\
Flat Layer   & 150            \\
Epochs         & 30        \\
Batch Size     & 800         \\
\hline
\hline
\end{tabular}
\caption{CNN architectures used for truth and reco cases.  \label{CNN-architecture}}
\end{table}

\bibliographystyle{JHEP}
{\small
\bibliography{main.bib}}

\end{document}